\newcommand{\be}{\begin{equation}}\newcommand{\ee}{\end{equation}}
\newcommand{\bea}{\begin{eqnarray}}\newcommand{\eea}{\end{eqnarray}}
\newcommand{\pa}{\partial}
\newcommand{\p}[1]{(\ref{#1})}
\newcommand{\lpr}{{\cal L}^{+4}}

\newcommand{\fracds}[2]{\displaystyle \frac{#1}{#2} }

\def\a{\alpha} \def\b{\beta} \def\e{\epsilon}

\def\G{\Gamma} 
\documentclass[12pt]{article}
\usepackage{amsfonts,colordvi}
\usepackage{amssymb}
\usepackage{amsfonts,amstext,latexsym,xspace}
\newcommand{\eq}{\begin{equation}}
\newcommand{\en}{\end{equation}}
\newcommand{\eqn}{\begin{eqnarray}}
\newcommand{\enn}{\end{eqnarray}}
\newcommand{\beq}{\begin{equation}}
\newcommand{\eeq}{\end{equation}}

\usepackage{amsmath}
\usepackage{amssymb}
\usepackage{amsfonts,amstext,latexsym,xspace}
\usepackage{epsfig}

\newcommand{\gd}[1][{}]{\delta_{#1}{}}

\renewcommand{\i}{\mathrm{i}}

\newcommand{\cX}{\mathcal{X}}
\newcommand{\cY}{\mathcal{Y}}

\newcommand{\cN}{\mathcal{N}}
\newcommand{\cI}{\mathcal{I}}

\begin{document}

\begin{titlepage}
\begin{center}
\begin{large}
\textbf{Harmonic Superspace, Minimal Unitary Representations   and  Quasiconformal Groups }
\end{large}
\vspace{1cm}
\\
\begin{large}
Murat G\"{u}naydin\footnote{murat@phys.psu.edu}
\end{large}
\\
\vspace{.35cm}
\emph{School of Natural Sciences \\
Institute for Advanced Study \\
Princeton, New Jersey 08540} \\
\vspace{.35cm} \begin{center} and \end{center}
 \emph{Physics Department \\
Pennsylvania State University\\
University Park, PA 16802, USA} \\
\vspace{.3cm}
\end{center}

\begin{abstract}
\noindent We show that there is a remarkable  connection
 between the harmonic
superspace (HSS) formulation of $N=2$, $d=4$ supersymmetric quaternionic K\"ahler sigma models that couple to $N=2$ supergravity and the minimal unitary representations of their
isometry groups. In particular, for $N=2$ sigma models with quaternionic
symmetric target spaces of the form $G/H\times SU(2)$ we establish
 a one-to-one mapping between the Killing potentials that generate the
  isometry group $G$  under Poisson brackets in the HSS formulation and
  the generators of the minimal unitary representation
  of $G$ obtained by quantization of its geometric realization as
   a quasiconformal group.
       Quasiconformal extensions  of  U-duality groups of four dimensional   $N=2$, $d=4$
       Maxwell-Einstein supergravity  theories (MESGT)  had been proposed as spectrum generating
  symmetry groups earlier. We  discuss some of the implications of our results, in particular, for the BPS
  black hole spectra of 4d $N=2$ MESGTs. 
  \end{abstract}
\end{titlepage}

\section{Introduction}
The target manifolds of $N=2$ supersymmetric $\sigma$ models  coupled  to supergravity in $d=4$
were shown to be quaternionic K\"ahler manifolds long time ago \cite{baggwitt}. Later the results
of \cite{baggwitt} were reformulated in harmonic superspace \cite{gios,Ivanov} as well as in projective
superspace \cite{N=2projective}. Some of these  theories arise as subsectors of the low energy effective
theories of type IIA or type IIB superstring compactified over a Calabi-Yau threefold. More
specifically,  type IIA (IIB) theories over  Calabi-Yau manifolds lead to $d=4, N=2$ supergravity
coupled to $h_{(1,1)} [ h_{(2,1)} ]$ vector multiplets and $(h_{(2,1)}+1) [(h_{(1,1)}+1)]$
hypermultiplets. Under further dimensional reduction  to three dimensions the vector moduli spaces
can be written as quaternionic K\"ahler manifolds ( c-map) \cite{GST1,cmap} and hence the moduli
spaces become products of two quaternionic K\"ahler manifolds in $d=3$.

In this paper we study the  symmetries of the $N=2$ supersymmetric $\sigma$ models with
quaternionic K\"ahler target manifolds that couple to supergravity in $d=4$ using the harmonic
superspace (HSS) approach and  show that there is  a remarkable mapping  between the realization of
the symmetries in the HSS formulation and the minimal unitary realizations of their isometry
groups. This mapping is made readily manifest  within the formulation of minimal unitary
realizations of noncompact simple groups obtained  by quantization of their geometric realizations
as quasiconformal groups \cite{GKN1,GKN2,GP1,GP2,GP3}. For $N=2$ supersymmetric $\sigma$ models the
relevant real forms are quaternionic.

 The plan of the paper is as follows. Section
   2 is devoted to a review of the formulation of  $N=2$ supersymmetric quaternionic K\"ahler $\sigma$ models in
     harmonic superspace and  implementation
   of their isometry groups through Killing potentials. In section 3 we review the quasiconformal
  realizations of noncompact groups that can be formulated simply and in full generality using the
   Freudenthal triple systems associated with them. The quasiconformal groups can all be defined
   as invariance groups of generalized
   lightcones  defined with respect to a quartic norm.
   In section 4 we review the
  construction of the minimal unitary representations of noncompact groups by
   quantization of their geometric realizations as quasiconformal groups.  In section 5 we show that for a
    quaternionic symmetric $N=2$ $d=4$ $\sigma$ model there is a precise mapping between the Killing 
potentials
     of its isometry
 group $G$ and the generators of the minimal unitary realization of $G$ given in sections 2 and 4, 
respectively.
  In section 6 we discuss some of the
 implications  of our results and open problems. In particular, we discuss the implications for the proposal that the quasiconformal extensions of
U-duality groups of  $d=4$ Maxwell-Einstein supergravity theories act as
spectrum generating symmetry groups.

\section{ $N=2$, $d=4$  Supersymmetric $\sigma$-models in Harmonic Superspace}

The target spaces of $N=2$ supersymmetric $\sigma$-models coupled to
 $N=2$ supergravity  are quaternionic K\"ahler manifolds \cite{baggwitt}. In this section
 we shall review the formulation of  these
$\sigma$-models in $N=2$, $d=4$, harmonic superspace \cite{bagger,gikos,gio} following closely
\cite{galogi}.  In the harmonic superspace approach  the  metric on a quaternionic target space is
determined by a  quaternionic potential ${\cal L}^{+4}$, which  plays the same role for
quaternionic K\"ahler manifolds as the K\"ahler potential does for K\"ahler manifolds.

The $N=2$ harmonic superspace action for the general $4n$-dimensional quaternionic $\sigma$-model
is given by \cite{galogi}
 \be S=\int
d\zeta^{-4}du \{ Q_\alpha^+D^{++}Q^{+\alpha} -q^+_i D^{++} q^{+i} + \lpr
(Q^+,q^+,u^-)\}.\label{action}\ee where the integration is over the analytic superspace coordinates
$\zeta, u^{\pm}_i$. The  $Q^+_\alpha(\zeta, u), \; \alpha =1,...,2n$ and the supergravity
hypermultiplet compensators $q^+_i(\zeta, u), \;(i=1,2)$ are analytic $N=2$ superfields. The   $
u^\pm_i,(i=1,2)$ are the $S^2= \frac{SU(2)_A}{U(1)}$ isospinor harmonics that satisfy \[
u^{+i}u^-_i=1
\] and $D^{++}$ is a supercovariant derivative with respect to harmonics with the property
\[ D^{++}u^-_i=u^+_i \]

We recall that the analytic subspace of $N=2$ harmonic superspace involves only half the Grassmann
variables with coordinates $\zeta^M$ and  $u^{\pm}_i $ 

 \be \zeta^M := \{ x^\mu_A , \;  \theta^{a +}
, \; \bar{\theta}^{\dot{a}+} \} \ee where \[ x^\mu_A := x^\mu -2i \theta^{(i} \sigma^\mu
\bar{\theta}^{j)} u_i^+ u_j^- \]
\[ \theta^{a +} := \theta^{a i} u_i^+ \]
\[ \bar{\theta}^{\dot{a}+} :=\bar{\theta}^{\dot{a} i} u_i^+ \]
\[ \theta^{(i} \sigma^\mu \bar{\theta}^{j)} u_i^+ u_j^- := \theta^{(ai} (\sigma^\mu)_{a\dot{a}}
\bar{\theta}^{\dot{a}j)} u_i^+ u_j^- \]
\[ \mu=0,1,2,3 ; \; a=1,2 ; \; \dot{a}=1,2 \]

This analytic subspace does not involve $U(1)$ charge $-1$ projections of the Grassmann variables
and is closed under $N=2$ supersymmetry transformations. Furthermore it is "real" with respect to
the conjugation $\widetilde{}$
\begin{eqnarray}
\widetilde{x}^\mu=x^\mu  \nonumber \\
\widetilde{\theta}^+ = \bar{\theta}^+ \\
\widetilde{\bar{\theta}}^+ = - \theta^+ \nonumber \\
\widetilde{u}^{i\pm} = u_i^{\pm}  \nonumber \\
\widetilde{u}_i^{\pm} - - u^{\i\pm}  \nonumber
\end{eqnarray}
which is a product of  complex conjugation and anti-podal map on the sphere $S^2$. For a complete
description of harmonic superspace we refer to the monograph  \cite{gios}.

The quaternionic potential $\lpr$ is a homogeneous function in $Q^+$  and $ q^+$ of degree two and
has  $U(1)$-charge $+4$.  It does not depend on $u^+$ and  is an arbitrary "real function"
otherwise, with the reality being  defined with respect to the involution $\widetilde{}$. For
simplicity we shall suppress the dependence of all the fields on the harmonic superspace
coordinates $\zeta^M$ and $u^{\pm}_i$.  

 As was first pointed out in \cite{go} and  later elaborated in  \cite{galogi,gios}
 the action \p{action} has a
remarkable analogy to the Hamiltonian mechanics with the harmonic derivative $D^{++}$ playing the
role of  time derivative. The superfields $Q^+$ and $ q^+$ correspond to phase space coordinates and the Poisson brackets
are given by \be \{f,g\}^{--} = {1\over 2}\Omega^{\alpha\beta}{\pa f\over \pa Q^{+\alpha}} {\pa
g\over \pa Q^{+\beta}}- {1\over 2}\epsilon^{ij}{\pa f\over\pa q^{+i}} {\pa g\over\pa q^{+j}},
\label{pb}\ee where $\Omega^{\alpha\beta}$ and $\epsilon^{ij}$ are the invariant antisymmetric
tensors of $Sp(2n)$ and $Sp(2)$ , respectively. The indices are raised and lowered by these tensors
\be Q^{+\alpha}= \Omega^{\alpha\beta}Q^+_\beta \ee \[ q^{+i}=\epsilon^{ij}q^+_j \] which
satisfy\footnote{ Note that the conventions of \cite{galogi} which we follow in this section differ
from those of \cite{GP3}.}

\be \Omega^{\alpha\beta}\Omega_{\beta\gamma} = \delta^{\alpha}_{\gamma} \ee \be \epsilon^{ij}
\epsilon_{jk}= \delta^i_k \ee
Because of this analogy and following \cite{galogi} we shall refer to the quaternionic potential
$\lpr$ as the Hamiltonian.

Isometries of the $\sigma$-model \p{action} are generated by Killing potentials $K^{++}_A(Q^+, q^+,
u^-)$ that obey the conservation law \be \pa^{++}K^{++}_A +\{K^{++}_A, \lpr\}^{--}=0\
\label{cons}\ee where $\partial_{++}$ is defined as
\[\partial_{++} = u^{+i} \frac{\partial}{\partial u^{-i}} \]
The Killing potentials form the Lie algebra of the isometry group \be \{K^{++}_A,
K^{++}_B\}^{--}=f_{AB}^{\ \ \ C}K^{++}_C\ .\label{Lie} \ee under the Poisson brackets \p{pb}.

The "Hamiltonians" of $N=2$ $\sigma$-models  coupled to N =2 supergravity with symmetric target
manifolds were given in \cite{galogi}. The quaternionic symmetric spaces , sometimes known as Wolf
spaces  \cite{wolf}, that are relevant to supergravity  are of the non-compact type. For each
simple Lie group  there is a unique non-compact quaternionic symmetric space.  A complete list of
these spaces is given below.
 \be\begin{array}{cccc} \fracds{SU(n,2)}{U(n) \times
Sp(2)}&\ \fracds{SO(n,4)}{SO(n)\times SU(2) \times Sp(2)}&  \ \fracds{USp(2n,2)}{Sp(2n)\times
Sp(2)} &
\fracds{G_{2(+2)}}{SU(2)\times Sp(2)}\\[3ex]
\fracds{F_{4(+4)}}{Sp(6)\times Sp(2)}&\ \fracds{E_{6(+2)}}{SU(6)\times Sp(2)}&\
\fracds{E_{7(-5)}}{SO(12)\times Sp(2)}& \fracds{E_{8(-24)}}{E_7 \times Sp(2)}.
\end{array}
\label{one} \ee which are all of the form  $G/H\times Sp(2)$ with $H \subset  Sp(2n)$.

For  a given quaternionic symmetric target space $G/H\times Sp(2)$ of $N=2$ $\sigma$ model
coordinatized by $Q^+$ and  $q^+$ , every generator $\G_A$ of $G$ maps to a function $K^{++}_A(Q^+, q^+,
u^-)$ such that the action of $K^{++}_A$ is given via the Poisson brackets \p{pb}. The authors of
\cite{galogi} showed that  the Hamiltonian $\lpr$ depends only on $Q^+$ and the combination
$q^+u^-:=q^{+i}u_i^-$, \be \lpr =\lpr (Q^+,(q^+u^-)) .\label{qu} \ee and can be written as \be \lpr
={P^{+4}(Q^+)\over (q^+u^-)^2} \label{formofl}\ee The fourth order polynomial $P^{+4}$ is given by
\be P^{+4}(Q^+)=\frac{1}{12} \; S_{\alpha\beta\gamma\delta}\;
Q^{+\alpha}Q^{+\beta}Q^{+\gamma}Q^{+\delta}\label{P^4} \ee where $S_{\alpha\beta\gamma\delta}$  is
a completely symmetric invariant tensor of $H$. In terms of  matrices $t^a_{\alpha\beta}, \; a=
1,2,.., dim(H)$ representing the action of the Lie algebra  $\mathfrak{h}$ of $H$ on $Q^{+\alpha}$
the invariant tensor reads as \cite{galogi} \be S_{\alpha\beta\gamma\delta}= h_{ab} t^a_{\ \alpha\beta}t^b_{\ \gamma \delta}
+\Omega_{\alpha\gamma}\Omega_{\beta\delta}+\Omega_{\alpha\delta}\Omega_{\beta\gamma} \ .\label{z12}\ee where $h_{ab}$ is the Killing metric of
$H$.

The Killing potentials that generate the isometry group $G$ are \cite{galogi}
 \be
\mathbf{Sp(2)}:\;\;\;\; K^{++}_{ij}=2(q^+_iq^+_j-u^-_iu^-_j\lpr), \label{su2curr} \ee
\be \mathbf{H}:\;\;\;\;
K^{++}_a=t_{a\alpha\beta}Q^{+\alpha}Q^{+\beta},\label{Hcurr} \ee
\be \mathbf{G/H\times
Sp(2)}:\;\;\;\;K^{++}_{i\alpha}=2q^+_iQ^+_\alpha - u^-_i(q^+u^-) \pa^-_\alpha\lpr\
,\label{cosetcurr} \ee
 where
\[ \pa^-_\alpha :=\frac{\partial}{\partial Q^{+\alpha}} \]  
\[ t_{a\a\b}=\Omega_{\beta\gamma}t_{a\alpha}^{\ \ \ \gamma} \]
 The $Sp(2)$ potentials $K^{++}_{ij}$ are conserved for an
arbitrary polynomial $P^{+4}(Q^+)$.
  $t^a$ are the representation matrices of the generators of $H$ acting on $Q^{+\alpha}$.  This implies  
that
the fourth order polynomial  $P^{+4}$  is proportional to the quadratic "Casimir function"
$h^{ab}K_a^{++}K_b^{++}$ of $H$. Furthermore , $P^{+4}$ can also be expressed in terms of the coset
Killing potentials, or the $Sp(2)$ Killing potentials  as follows \cite{galogi} : \be
P^{+4}=-\frac{1}{16}\e^{ij}\Omega^{\alpha\beta}K^{++}_{i\a} K^{++}_{j\b}=-\frac{1}{8}K^{++ij}K^{++}_{ij}\ .\ee

\section{Freudenthal Triple Systems and Quasiconformal Group Actions}
Lie algebra $\mathfrak{g}$ of every simple Lie group $G$ admits a  5-graded decomposition of the form
\begin{eqnarray}
\mathfrak{g} = \mathfrak{g}^{-2} \oplus \mathfrak{g}^{-1} \oplus
       \mathfrak{g}^0 \oplus \mathfrak{g}^{+1}
      \oplus \mathfrak{g}^{+2} \,.
\end{eqnarray}
such that grade $\pm 2$ subspaces are one dimensional.\footnote{ Of course for
$sl(2)$ this 5-grading degenerates to a 3-grading.} The grade zero algebra
$\mathfrak{g}^0$ has the form \be \mathfrak{g}^0 = \mathfrak{h} \oplus \Delta
\ee where $\Delta$ is the generator that determines  the 5-grading. The grade
$\pm 2$ generators and $\Delta$ generate a distinguished  $sl(2)$ subalgebra of
$\mathfrak{g}$.
 We shall denote the subgroup generated by  $\mathfrak{h}$ as $H$.
A simple Lie algebra with such a 5-graded  decomposition  can always be
constructed over a Freudenthal triple system $\mathcal{F}$
\cite{Freudenthal,kansko}. Freudenthal introduced these triple systems   in his
study of the meta\-symplectic geometries associated with exceptional groups
\cite{Freudenthal}. A Freudenthal triple system (FTS) is a vector space
$\mathcal{F}$ with a trilinear product $(X,Y,Z)$ and a skew symmetric bilinear
form $<X,Y>$ that satisfy \footnote{It should be noted that the triple product
can be modified by terms involving the symplectic invariant, such as $\langle
X,Y \rangle Z$.  The  choice given above was made in \cite{GKN1}.}:
\begin{eqnarray}
(X,Y,Z) &=&(Y,X,Z) +2\,\langle X,Y \rangle Z \,,\nonumber \\
(X,Y,Z) &=& (Z,Y,X) -2\,\langle X,Z \rangle Y \,,\nonumber \\
\langle (X,Y,Z), W \rangle  &=& \langle (X,W,Z),Y \rangle
                                -2\,\langle X, Z \rangle \langle Y ,W \rangle \,,\nonumber \\
(X,Y,(V,W,Z)) &=& (V,W,(X,Y,Z)
                             +((X,Y,V),W,Z) \nonumber \\
                          && {}+ (V,(Y,X,W),Z)  \,.\label{ftp56-rel}\end{eqnarray}
A  quartic invariant $\mathcal{I}_4$ can be defined over the FTS $\mathcal{F}$ using
the  triple product and the bilinear form  as
\begin{eqnarray}
\mathcal{I}_4(X) := \frac{1}{48}\langle (X,X,X),X \rangle \label{e7-invariant}
\end{eqnarray}
which is invariant under the automorphism group $H=Aut(\mathcal{F})$   of $\mathcal{F}$.

In the corresponding construction of  $\mathfrak{g}$ over $\mathcal{F}$,    the
generators of grade $\pm 1$ subspaces of $\mathfrak{g}$ are labelled by the
elements of $\mathcal{F}$ and all the commutation relations are expressed in
terms of the triple product $(X,Y,Z)$ \cite{kansko}. Following \cite{GKN1} let us denote
the  Lie algebra generators belonging to grade $+1$ and grade $-1$ subspaces as
$U_A$ and $\widetilde U_A$, respectively, where $A \in \mathcal{F}$. The five
grading now reads as
\[ \mathfrak{g}= \widetilde{K}_{AB} \oplus \widetilde{U}_A \oplus S_{AB} \oplus U_A \oplus K_{AB} \]
where $A,B \in  \mathcal{F}$. The symplectic trace of  $S_{AB}$ is the
generator $\Delta$ that determines the five grading \cite{mg92}. Since  they
are one dimensional the grade $\pm 2$  generators $K_{AB}$ and $\widetilde K_{AB}$
labeled by two elements can be written as
\begin{equation}
  K_{AB} = \left<A,B\right> K \qquad\quad
  \widetilde{K}_{AB} =\left<A,B\right> \widetilde{K}
\end{equation}
Hence we have,
\[ \mathfrak{g} = \widetilde{K} \oplus \widetilde{U}_A \oplus S_{AB} \oplus U_A \oplus K \]
Commutation relations among these generators  in terms of
the  triple product of $\mathcal{F}$ was given in \cite{GKN1} following earlier references \cite{Freudenthal,kansko}.

As was shown in \cite{GKN1} one can realize the Lie algebra $\mathfrak{g}$  as
a quasiconformal Lie algebra over a vector space $\mathcal{Q}$ whose
coordinates $\cX$ are labeled by a pair $\left(X,x\right)$, where $X \in
\mathcal{F}$ and $x$ is an extra single variable as follows :

\begin{equation}
\begin{split}
  \begin{aligned}
      K\left(X\right) &= 0 \\
      K\left(x\right) &= 2\, a
  \end{aligned}
  & \quad
  \begin{aligned}
     U_A \left(X\right) &= A \\
     U_A\left(x\right) &= \left< A, X\right>
  \end{aligned}
   \quad
   \begin{aligned}
      S_{AB}\left(X\right) &= \left( A, B, X\right) \\
      S_{AB}\left(x\right) &= 2 \left< A, B\right> x
   \end{aligned}
 \\
 &\begin{aligned}
    \widetilde{U}_A\left(X\right) &= \frac{1}{2} \left(X, A, X\right) - A x \\
    \widetilde{U}_A\left(x\right) &= -\frac{1}{6} \left< \left(X, X, X\right), A \right> + \left< X, A\right> x
 \end{aligned}
 \\
 &\begin{aligned}
    \widetilde{K}\left(X\right) &= -\frac{1}{6} \,  \left(X,X,X\right) +  X x \\
    \widetilde{K}\left(x\right) &= \frac{1}{6} \,  \left< \left(X, X, X\right), X \right> + 2\,  \, x^2
 \end{aligned}
\end{split}
\end{equation}
The symplectic traceless components of $S_{AB}$ generate the  automorphism group $H$ of the FTS
$\mathcal{F}$ and the trace part ($\Delta$) is the generator  that determines  the 5-grading.

One defines a quartic norm over the space $\mathcal{Q}$ as
\be
\cN_4(\cX) := \cI_4(X) - x^2
\ee
and the "distance" between any two points $\cX=(X,x)$ and  $\cY=(Y,y) $ in $\mathcal{Q}$ as
\eq
d(\cX,\cY):= \cN_4(\gd(\cX,\cY)
\en
where  $\gd(\cX,\cY)$ is the "symplectic" difference vector of two  vectors $\cX $ and $\cY$ :
\[  \gd(\cX,\cY)= (X-Y,x-y+\langle X, Y \rangle )\]

The invariance of $d(\cX,\cY)$ under the action of the automorphism group of $\mathcal{F}$ and
"translations" $U_A$ and $K$ is manifest. The generator $\Delta$ simply rescales $d(\cX,\cY)$,
while  under the  action of the negative grade generators one finds that $d(\cX,\cY)$ gets
multiplied by functions linear in $\cX$ and $\cY$. Hence the quasiconformal group action preserves
light-like separations
\[
  d(\cX,\cY)=0
\]
defined by the quartic norm. This is the reason why  the above geometric action
of $G$ was called quasiconformal in \cite{GKN1}.

Here we should stress an important point. The construction  of a simple Lie
algebra $\mathfrak{g}$  over a FTS
 $\mathcal{F}$ extends in a straightforward manner to the complex Lie algebra $\mathfrak{g}(\mathbb{C})$
 by complexifying $\mathcal{F}$. Then the above realization of the quasiconformal action of $G$ extends to
 a quasiconformal action of $G(\mathbb{C})$. One can then obtain  quasiconformal realizations
 of  different  real forms of $G$ by appropriate restriction of the complex $G(\mathbb{C})$.

\section{ Minimal unitary representations
of non-compact groups from their quasiconformal realizations}

In this section we shall review the unified construction of  minimal unitary representations
 of noncompact groups obtained by quantization of their
geometric realizations as quasiconformal groups following \cite{GP3} which generalizes earlier
results of \cite{GKN2,GP1,GP2}.
Consider the 5-graded decomposition of  the Lie algebra
$\mathfrak{g}$ of a noncompact group $G$

\begin{equation}
\mathfrak{g}=  \mathfrak{g}^{-2}  \oplus  \mathfrak{g}^{-1}  \oplus \left(  \mathfrak{h}  \oplus \Delta 
\right)
  \oplus  \mathfrak{g}^{+1}  \oplus  \mathfrak{g}^{+2} \nonumber
\end{equation}
\be \mathfrak{g} = E \oplus E^\alpha \oplus ( J^a + \Delta ) \oplus F^\alpha \oplus F \ee
where $\Delta$ is the generator that determines the 5-grading.
Generators $J^a$  of  $\mathfrak{h}$ satisfy
\begin{subequations}\label{eq:alg}
\begin{equation}
   \left[ J^a \,, J^b \right] = {f^{ab}}_c J^c
\end{equation}
where $a,b,...=1,...D=dim(H)$.  Let $\rho$ denote the symplectic
representation by which $\mathfrak{h}$ acts on $\mathfrak{g}^{\pm
1}$
\begin{equation}
   \left[ J^a \,, E^{\alpha} \right] = {\left(\lambda^{a}\right)^\alpha}_\beta E^\beta
\qquad
   \left[ J^a \,, F^{\alpha} \right] =  {\left(\lambda^{a}\right)^\alpha}_\beta F^\beta
\end{equation}
where $E^\alpha$, $ \alpha, \beta, ..= 1,..,N= \dim (\rho)$ are
generators that span the subspace $\mathfrak{g}^{-1}$
\begin{equation}
   \left[ E^\alpha \,, E^\beta  \right] = 2 \Omega^{\alpha\beta} E
\end{equation}
and $F^\alpha$ are generators that span $\mathfrak{g}^{+1}$
\begin{equation}
   \left[ F^\alpha \,, F^\beta  \right] = 2 \Omega^{\alpha\beta} F
\end{equation}
 $\Omega^{\alpha\beta}$ is the symplectic invariant ``metric'' of
the representation $\rho$.  The positive  (negative)   grade generators form a
Heisenberg subalgebra since
\begin{equation}
  \left[E^{\alpha}, E \right] = 0
\end{equation}
with the grade +2 (-2) generator $F$ ($E$)   acting as its central charge.
 The remaining nonvanishing commutation relations of $\mathfrak{g}$ are
\begin{equation}
\begin{aligned}
  F^\alpha &= \left[ E^\alpha \,, F \right] \cr
  E^\alpha &= \left[ E \,, F^\alpha \right] \cr
  \left[E^{\alpha} , F^{\beta}\right] &= - \Omega^{\alpha\beta} \Delta + \epsilon \lambda_a^{\alpha\beta} 
J^a
\end{aligned}
\qquad
\quad
\begin{aligned}
\left[\Delta, E^{\alpha} \right] &= - E^{\alpha} \cr
\left[\Delta, F^{\alpha} \right] &= F^{\alpha} \cr
\left[\Delta, E\right] &= -2 E \cr
\left[\Delta, F \right] &= 2 F
\end{aligned}
\end{equation}
\end{subequations}
where
$\epsilon$ is a  constant parameter whose value depends on the Lie algebra $\mathfrak{g}$.

 In the unified minimal unitary realization of noncompact groups \cite{GP3}, negative grade
   generators are expressed as bilinears of  bosonic oscillators
$\xi^\alpha$ satisfying the canonical commutation relations
\begin{equation}
   \left[ \xi^\alpha \,, \xi^\beta \right] = \Omega^{\alpha\beta}
\end{equation}
and an extra coordinate $y$ ,corresponding to the singlet in their quasiconformal realization
\footnote{Here let us  emphasize that we are thereby realizing the Heisenberg algebra $ \mathfrak{g}^{-2}  \oplus  \mathfrak{g}^{-1} $ in terms of coordinate and momentum operators $\xi^\alpha$ , modulo a scale coordinate $y$ which determines the central charge $E=\frac{1}{2}y^2$ . This is what we mean by {\it quantization} of the geometric action of the quasiconformal group.}

\begin{equation}
  E = \frac{1}{2} y^2 \qquad E^\alpha = y \,\xi^\alpha \qquad
 J^a =  - \frac{1}{2}  {\lambda^a}_{\alpha\beta} \xi^\alpha \xi^\beta
\end{equation}

The quadratic Casimir operator of the Lie algebra $\mathfrak{h}$ is
\begin{equation}
 \mathcal{C}_2\left(\mathfrak{h}\right) = \eta_{ab} J^a J^b
\end{equation}
where $\eta_{ab}$ is the Killing metric of the subgroup  $H$ , which is isomorphic to the
 automorphism group of the underlying FTS $\mathcal{F}$. The quadratic Casimir $C_2(\mathfrak{h})$
 is equal to the quartic invariant of $H$ in the representation $\rho$ modulo an additive constant
 that depends on the normal ordering chosen, namely
\be I_4(\xi^\alpha) = S_{\alpha\beta\gamma\delta} \xi^\alpha \xi^{\beta} \xi^{\gamma} \xi^{\delta}
=  C_2(\mathfrak{h}) + \mathfrak{c}\ee
where $\mathfrak{c}$ is a  constant and \[ S_{\alpha\beta\gamma\delta} := \lambda_{a(\alpha\beta} 
\lambda^a_{\gamma\delta)}\]

The grade +2 generator $F$  has the general form
\begin{equation} \label{eq:exprF}
   F = \frac{1}{2} p^2 + \frac{\kappa  \left( \mathcal{C}_2(\mathfrak{h}) + \mathfrak{C} \right)}{y^2}
\end{equation}
where $p$ is the momentum conjugate to the singlet coordinate $y$
\begin{equation}
[y,p]=i
\end{equation}
and $\kappa$ and $\mathfrak{C}$ are some constants depending on the Lie algebra $\mathfrak{g}$.
The grade $+1$ generators are then given by
\begin{equation}
  F^\alpha = \left[E^{\alpha}, F\right]= i p \, \xi^\alpha + \kappa y^{-1} \left[ \xi^\alpha \,, \mathcal
{C}_2
  \right]
  \end{equation}
  For simple or Abelian $H$ they take the form \cite{GP3}
  \begin{equation}
 F^\alpha= ip \, \xi^\alpha -\kappa y^{-1} \left[ 2 \, {\left(\lambda^{a}\right)^{\alpha}}_\beta \xi^\beta 
J_a
  + \, C_\rho \,\xi^\alpha \right]
\end{equation}
where $C_\rho$ is the eigenvalue of the second order Casimir of $H$
in the representation $\rho$ and one finds

\begin{equation}
  \left[ E^\alpha \,, F^\beta \right] = - \Delta \Omega^{\alpha\beta} - 6 \kappa \left(\lambda^a\right)^
{\alpha\beta} J_a
\end{equation}
where $\Delta = - \frac{i}{2} \left( yp+py\right)$. \footnote{ In this section  we follow the
conventions of \cite{GP3}.  The indices $\alpha, \beta,..$ are raised and lowered with the
antisymmetric symplectic metric $\Omega^{\alpha\beta}=-\Omega^{\beta\alpha}$ that satisfies
$\Omega^{\alpha\beta}\Omega_{\gamma\beta}=\delta^{\alpha}_{\beta} $ and $V^\alpha =
\Omega^{\alpha\beta} V_\beta $, and $V_\alpha =V^\beta \Omega_{\beta\alpha} $.}

Using the results of \cite{bg} one can give a unified realization of all simple Lie algebras
in terms of the underlying FTS's $\mathcal{F}$ \cite{GP4}. In the most general case one finds that the
commutator of $ E^\alpha $ and $ F^\beta $ has the same form as above , namely \cite{GP4}
\[
  \left[ E^\alpha \,, F^\beta \right] = - \Delta \Omega^{\alpha\beta} -
  \epsilon \left(\lambda^a\right)^{\alpha\beta} J_a
\]
where $\epsilon$ is a constant and
$\left(\lambda^a\right)^{\alpha\beta} $ are the matrices of the Lie algebra of  automorphism group $H$ of
the underlying FTS $\mathcal{F}$.

For simple $H$ the quadratic Casimir operator of the Lie algebra $\mathfrak{g}$ is given by \cite{GP3}

\begin{equation}
  \mathcal{C}_2 \left( \mathfrak{g} \right) = J^a J_a + \frac{2 \, C_\rho}{N+1} \left( \frac{1}{2}\, 
\Delta^2 + E F + F E \right) - \frac{ C_\rho}{N+1}
  \Omega_{\alpha\beta} \left( E^\alpha F^\beta + F^\beta E^\alpha \right)
\end{equation}
Furthermore one finds that the quadratic Casimir of $sl(2)$ and the contribution of the coset
generators $F^\alpha$ and $E^\beta$ to  $\mathcal{C}_2 \left( \mathfrak{g} \right)$ can all be expressed in 
terms of the quadratic Casimir $J^aJ_a$  of $H$ :

\begin{equation}
\begin{split}
   \frac{1}{2}\, \Delta^2 + E F + F E &= \kappa\left( J^a J_a + \mathfrak{C} \right) - \frac{3}{8} \\
   \Omega_{\alpha\beta}\left( E^\alpha F^\beta + F^\beta E^\alpha \right) &= 8 \, \kappa J^a J_a + \frac{N}
{2} + \kappa C_\rho N
\end{split}
\end{equation}
 and the quadratic Casimir of $\mathfrak{g}$ reduces to a c-number\cite{GP3}
\begin{equation}
\mathcal{C}_2 \left( \mathfrak{g} \right) =  \mathfrak{C} \left( \frac{8 \kappa C_\rho}{N+1} - 1 \right) - 
\frac{3}{4} \, \frac{C_\rho}{N+1} - \frac{N}{2} \, \frac{C_\rho}{N+1} - \frac{\kappa C_\rho^2 N}{N+1}
\end{equation}
as required by irreducibility of the minimal representation.  This is a general  phenomenon for all minimal unitary realizations of 
simple groups $G$ \cite{GKN2,GP1,GP2,GP3,GP4}.

 \section{Mapping  between Killing Potentials in HSS and generators of minimal unitary representations of  isometry groups of $\sigma$-models}

 To establish a precise mapping between the Killing potentials of the isometry group $G$  of the sigma 
model in harmonic superspace and the generators of the minimal unitary realization
 of $G$ we shall rewrite the Killing potentials in an $SU(2)_A$ invariant manner by contracting the 
generators given in section 4 with the spherical harmonics $u^{+i}$ and $u^{-i}$. First let us define \footnote{ The $w_c$ and $\frac{p_c}{w_c}$ are labelled as fields $w$ and $N^{++}$ 
and  interpreted geometrically as the central charge coordinates $Z^0$ and $Z^{++}$ in \cite{gios}.} 
 \be
\sqrt{2} q^{+i} u_i^- := w_c  \ee
 \be
 \sqrt{2} q^{+i} u_i^+ := p_c
 \ee
 The poisson brackets of $q^{+i}$
 \be \{q^{+i},q^{+j}\} = -\frac{1}{2} \epsilon^{ij} \ee
 imply that
 \be \{w_c,p_c\}= -1 \ee
 Under the conjugation $\widetilde{}$ we have \[ \widetilde{\widetilde{q^{+i}}} = - q^{+i} \]
 \[ \widetilde{\widetilde{u^{\pm}_i }} = - u_i^{\pm} \]
 which imply \be \widetilde{\widetilde{ w_c}} = w_c \ee
 \be \widetilde{\widetilde{p_c}} = p_c \ee
 The Hamiltonian  can then be written as \be \lpr ={2P^{+4}(Q^+)\over w_c^2}
\label{formofl}\ee

The $SU(2)_A$ invariant Killing potentials that generate the isometry group $G$ are then

 \be
\mathbf{Sp(2)}:\;\;\; S^{++}:= K^{++}_{ij}u^{+i}u^{+j} = p_c^2 - \frac{2P^{+4}(Q^+)}{ w_c^2},
\label{su2curr2} \ee
\be \;\;\;\;\;\;\;\;\;\;\;\;\;\;\;\; S^0= K^{++}_{ij} ( u^{+i}u^{-j} + u^{+j} u^{-i} )= w_c p_c +p_c w_c \ee
\be \;\;\;\;\; S^{--}= K^{++}_{ij} u^{-i}u^{-j} = w_c^2 \ee

\be \mathbf{H}:\;\;\;\;\;\;\;\;\;\;\;\; K^{++}_a=t_{a\alpha\beta}Q^{+\alpha}Q^{+\beta},\label{Hcurr} \ee

\be \mathbf{G/H\times Sp(2)}:\;\;\;\;  K^+_{\alpha}:= K^{++}_{i\alpha} u^{+i}= -\sqrt{2} \{ p_c Q^+_\alpha -
\frac{1}{w_c} \pa^-_\alpha  P^{+4}(Q^+)  \} ,\label{cosetcurr2} \ee
\be K^-_{\alpha}:= K^{++}_{i\alpha} u^{-i}= -\sqrt{2} w_c Q_{\alpha}^+ \ee

 Comparing the above Killing potentials of the isometry group $G$ with the generators of the minimal 
unitary realization of $G$ given in section 3 we have the following one-to-one correspondence between the elements of 
harmonic superspace (HSS) and those of  minimal unitary realizations (MINREP)
 \\
\begin{center}
  \begin{tabular}{|| c | c ||}
\hline
\hline
  HSS  ¥ & ¥MINREP  \\ \hline
   $w_c$ & $y$ \\
 & \\ \hline
  $ p_c $ ¥ & ¥$ p $ \\
 & \\ \hline
    $\{ \;,\;\}$¥ & $i[\;,\;]$¥ \\
 & \\ \hline
    $Q^{+\alpha}$¥ & ¥$\xi^{\alpha}$ \\
 & \\ \hline
    $P^{+4}(Q^+)$¥ & ¥$I_4(\xi)$ \\
 & \\  \hline
   $ K^{a++}=t^a_{\alpha\beta} Q^{+\alpha} Q^{+\beta}$ ¥ & ¥$J^a =\lambda^a_{\alpha\beta} \xi^{\alpha}\xi^
{\beta}$ \\  
 & \\  \hline
   $K_\alpha^+=K^{++}_{i\alpha} u^{+i}$ ¥ & ¥$F^\alpha$ \\
 & \\  \hline
  $ K_\alpha^-=K^{++}_{i\alpha} u^{-i}$ ¥ & ¥$E^{\alpha}$ \\
 & \\  \hline
    \hline
  \end{tabular}
\end{center}

The Poisson brackets (PB) $ \{ , \}$ in HSS formulation go over to $i$ times the commutator 
$ [ , ]$ in the minimal unitary realization and the 
classical harmonic superfields  $w_c,p_c$  , that are canonically conjugate under PB map to the canonically 
conjugate coordinate and momentum operators $y,p$. Similarly, the harmonic superfields $Q^{+\alpha}$ that 
form $n$ conjugate pairs under Poisson brackets go over to the oscillators $\xi^\alpha$. 
 This introduces a normal ordering ambiguity in the quantum version of the quartic invariant.
 Thus  the classical expression relating the quartic invariant polynomial $P^{+4}$ to the quadratic Casimir 
function in HSS  differs from the expression relating the quartic invariant $I_4$ to the quadratic Casimir
  of $H$ by an additive c-number depending on the ordering chosen. The consistent choices for the
  quadratic Casimir and  corresponding c-numbers for all  noncompact groups , whose quotients with respect to their maximal compact subgroup are quaternionic symmetric,  can 
be found in \cite{GP1,GP2,GP3}.

  The   mapping between HSS and MINREP extends also to the equations relating the quadratic Casimir of $\mathfrak{h}$ to the 
  quadratic Casimir of $\mathfrak{sp}(2)$ and to the contribution of the coset generators $G/H\times Sp(2)$ to 
the quadratic Casimir of $\mathfrak{g}$
modulo some additive constants due to normal ordering.

Of course on the MINREP side we are working with simple quantum mechanical coordinates and momenta, while 
in HSS the corresponding quantities are classical harmonic analytic superfields. The easiest
way to make more concrete the above mapping is to reduce the $4d$ $N=2$ $\sigma$ model to one dimension 
and quantize it to get a supersymmetric quantum mechanics ( with 8 superscharges). What the above mapping 
implies is that the spectrum of the corresponding quantum mechanics must furnish a minimal unitary 
representation of the isometry group , which is fully supersymmetric , since the supersymmetry generators commute with 
the isometry group.

\section{ Discussion}

We find the correspondence between the formulation of  $N=2$ , $d=4$ quaternionic K\"ahler $\sigma$ models in HSS and the minimal unitary 
realizations of their isometry groups established above quite remarkable. We will discuss some of the 
implications of this correspondence and  open problems, that will be the subjects of separate investigations.

It is important to extend the correspondence between the minimal unitary representation of the isometry 
group  and the classical $N=2$ $, d=4$ quaternionic K\"ahler $\sigma$ model to its quantum theory in HSS. 
There is a subtle issue regarding the quantum implementation of the conjugation $\widetilde{}$ with respect 
to which the harmonic derivative $D^{++}$ is real.  This extension to the quantum theory and resolution of  the subtle issues should be easier  if one reduces the quaternionic K\"ahler $N=2$ $\sigma$ model to two dimensions or to 
quantum mechanics with eight supersymmetries \cite{d1d2}. For the following discussion we shall assume that there is no 
obstruction to extending the mapping to the quantum theory.

The correspondence established for symmetric space theories implies that the {\it fundamental spectra } of the 
quantum $N=2$ , quaternionic K\"ahler $\sigma$ models in $d=4$  and their lower dimensional counterparts must fit  into the 
minimal unitary representations of their isometry groups. 
By the fundamental spectra we mean the well-defined states created by the action of  harmonic analytic superfields at a 
given point in analytic superspace with coordinates $\zeta^M$ on the vacuum of the theory. From the mapping 
above we expect that the states created by the purely bosonic components  of the analytic superfields will 
fit into the minimal unitary representation of the corresponding isometry group. Since the analytic 
superfields are unconstrained, the bosonic spectrum extends to an $N=2$ supersymmetric spectrum ( 8 supercharges) by the 
action of the fermionic components of the superfields. 
 
Now the minimal unitary representations are the analogs of the singleton representations of symplectic 
groups $Sp(2n, \mathbb{R})$.  The singleton realizations of  $\mathfrak{sp}(2n, \mathbb{R})
$  are free field realizations , i.e. their generators can be written as bilinears of bosonic oscillators . 
As a consequence the tensoring procedure becomes simple and straightforward for the symplectic groups \cite{singleton}. 
However , for other groups the minimal unitary realization is "interacting" and the corresponding 
generators are nonlinear in terms of the oscillators. This makes the tensoring problem highly nontrivial.
The tensoring of Fock spaces of  free bosons in the case of  $Sp(2n,\mathbb{R})$ will go over to  tensoring of corresponding minimal unitary representations for other noncompact groups.  For the quantum $N=2$ quaternionic K\"ahler $\sigma$ models  one then has to  
 tensor  the fundamental supersymmetric spectra with each other repeatedly.
 By an abuse of 
terminology we shall refer to the resulting spectra as  "perturbative" spectra in quantum HSS.
 The "nonperturbative" spectra in quantum HSS will , in general, contain states that do not form full $N=2$ 
supermultiplets.

The fundamental spectrum is  generated by the action of  analytic harmonic superfields involving an 
infinite number of auxiliary fields. Once the auxiliary fields are eliminated the dynamical components of 
the superfields become complicated nonlinear functions of the physical bosonic and fermionic fields. 
Therefore the fundamental spectrum in HSS correspond to states created by some complicated nonlinear functions of the physical 
fields in general. Hence the "fundamental spectrum" is in general not the simple Fock space of free bosons 
and fermions.  
 
 Since the HSS formulation extends to all $N=2$ supersymmetric $\sigma$ models in $d=4$, we expect the 
fundamental spectra of all $N=2$,  $\sigma$ models with nontrivial 
isometry groups to form minimal unitary representations of their isometry groups. An important class of 
$N=4$, $\sigma$ models in $d=3$ ,whose scalar manifolds are not homogeneous , but have interesting  
isometry groups can be obtained by dimensional reduction ( CR-map = C-map times R-map ) from unified $N=2$ 
Maxwell-Einstein  supergravity theories in $d=5$ \cite{GST1,GZ05}. These unified $d=5$, $N=2$ MESGTs with simple 
U-duality groups  belong to three infinite families plus a sporadic one. The sporadic theory and the lowest 
members of the three infinite families are the magical MESGTs whose scalar manifolds are symmetric spaces 
\cite{GST2} in 5, 4 and 3 dimensions. The scalar manifolds of the other unified theories in $d=5$ are 
neither symmetric nor homogeneous \cite{GZ05}. The resulting three dimensional $N=4$, quaternionic K\"ahler $\sigma$ models can then be 
lifted up to  four dimensional $N=2$ supersymmetric $\sigma$ models, with a rich family of interesting isometry groups. 

The $N=2$ , $d=4$ MESGT's  lead to $N=4$ , $d=3$ supersymmetric $
\sigma$ models with quaternionic K\"ahler manifolds $\mathcal{M}_3$ under dimensional reduction on a spacelike circle (C-
map).   On the other hand the stationary black hole solutions of $N=2$ MESGTs can be reduced to three 
Euclidean dimensions on a timelike circle. The resulting theory is $d=3$ Euclidean gravity coupled to 
a para-quaternionic K\"ahler manifold $\mathcal{M}^*_3$ \cite{GNPW1,GNPW2}.
  For radially symmetric stationary (supersymmetric) black holes the attractor equations become equivalent to (supersymmetric) geodesic motion on $
\mathcal{M}^*_3$ \cite{GibBre}. The radial quantization of BPS black hole solutions can then be implemented by replacing functions on  
classical 
phase space on $\mathcal{M}^*_3$ by square integrable functions on $\mathcal{M}_3$ \cite{GNPW1,GNPW2}.
Furthermore, the $8n$ dimensional general phase space is reduced to $4n+2$ dimensional subspace after imposing the BPS conditions, which can be identified with the twistor space of $\mathcal{M}_3$ \cite{GNPW2,NPV}. 
For very special symmetric quaternionic K\"ahler manifolds that are obtained from $d=5$ , $N=2$ MESGTs by dimensional reduction to three dimensions the corresponding manifolds are of the form
\be
\mathcal{M}_3 = \frac{QCon(J)}{Konf(J) \times SU(2)} 
\ee
\be 
 \mathcal{M}_3^* = \frac{QCon(J)}{Conf(J) \times Sl(2,\mathbb{R})} 
 \ee
 where $QConf(J)$ and $Conf(J)$ denote the quasiconformal and conformal groups of the Jordan algebras $J$ of degree three that define the corresponding five dimensional theory, respectively. $Konf(J)$ refers to the compact form of the conformal group $Conf(J)$ of the Jordan algebra $J$.  
 $Conf(J)$ has been proposed as a spectrum generating symmetry group of the 5d, $N=2$ MESGT 
 \cite{fergun,GKN1,MG} and $QConf(J)$ has been proposed as the spectrum generating symmetry 
 group  of 4d, $N=2$ MESGT defined by $J$ \cite{GKN1,MG}.
 The twistor space on $\mathcal{M}_3$ is simply \[ \frac{QConf(J)}{Konf(J)\times U(1)} \] 
 whose K\"ahler potential is given by the distance function that defines the quartic light-cone \cite{gnppw,GNPW2}.
 Hence  the BPS Hilbert space must form a unitary representation of   $G_3$ induced by the geometric realization of $G_3$ as a quasiconformal group \cite{GNPW2}. The unitary representations that arise this way belong to the quaternionic discrete series and are not the minimal unitary representations whose Gelfand-Kirillov dimensions are much smaller. Since the BPS states correspond to four supercharges this result is not surprising. However, the results presented above imply that the $N=4$, $d=3$ symmetric quaternionic K\"ahler $\sigma$ models have {\it fundamental spectra} which preserve all the supersymmetries.
 Whether there exist fully supersymmetric  black hole solutions belonging to the fundamental spectra in these theories is currently under investigation \cite{GNPW2}. \\
 
 {\bf Acknowledgements:} The main results of this paper were obtained  during the spring of 2006. A summary of these results was then  circulated ,in June 2006, among  a small group of colleagues. I would like to thank Andy Neitzke , Boris Pioline and Andrew Waldron for helpful  comments on this summary and  discussions. 
I   would also like to thank Juan Maldacena and Oleksandr Pavlyk for illuminating  discussions.  This work 
was supported in part by the National Science Foundation under grant number PHY-0555605. Any opinions, 
findings and conclusions or recommendations expressed in this material are those of the authors and do not 
necessarily reflect the views of the National Science Foundation. The support of Monell Foundation  during 
my stay at IAS, Princeton, is gratefully acknowledged.

\end{document}